\documentstyle[floats,aps,epsf,prl]{revtex}

\def \be{\begin{displaymath}}
\def \ee{\end{displaymath}}              

\def \ben{ \begin{equation} }
\def \een{ \end{equation}   }            

\def \bea{\begin{eqnarray*}}             
\def \eea{\end{eqnarray*}}

\def \bean{\begin{eqnarray}}             
\def \eean{\end{eqnarray}}

\def \nn{\nonumber}

\def \Ref#1{(\ref{#1})}
\def \fie {\varphi}

\def \e{ {\rm e}}

\def \inv{ ^{-1} }

\def \invb#1 { \frac{1}{#1} }

\def \av#1{ {\left\langle #1 \right\rangle} }

\def \fr#1#2{ \frac{#1}{#2} }

\def \gapprox{\mbox{\raisebox{-4pt}{$\,\buildrel>\over\sim\,$}}}

\def \angst{ \mbox{{\it{\AA}}}}

\begin{document}
\draft 
\twocolumn[\hsize\textwidth\columnwidth\hsize\csname @twocolumnfalse\endcsname

\title{Electron scattering in multi-wall carbon-nanotubes}

\author{Rochus Klesse}

\address{ Universit\"at zu K\"oln, Institut f\"ur Theoretische Physik,
     Z\"ulpicher Str. 77, D-50937 K\"oln, Germany.}

\date{June 14, 2002}

\maketitle

\begin{abstract}
We analyze two scattering mechanisms that
might cause intrinsic electronic resistivity in  multi-wall carbon
nanotubes: scattering by dopant impurities, and scattering by inter-tube
electron-electron interaction. We find that for typically doped
multi-wall tubes backward scattering at dopants is by far the dominating
effect.
\end{abstract}

\pacs{PACS}]

\section{Introduction}

Carbon nanotubes \cite{iijima} appear in a rich variety of
size and molecular structure. Moreover, they can self-arrange in well-defined
secondary structures, like multi-wall tubes or bundles of closely packed
single-wall tubes. This provides electronic systems ranging from
strictly one dimensional metals and semi-conductors, up to quasi
two-dimensional, graphite-like systems. Single-wall nanotubes with 
diameters of order $nm$ behave even at room temperature as strictly 
one-dimensional electronic systems. 
Multi-wall tubes, on the other hand, have typically rather large diameters
of several tenths of nanometers, and therefore exhibit less distinct
one-dimensional features.

While it is established that the physics of a conducting single-wall 
tube can be described by a four-channel Luttinger liquid
\cite{general,bockrath}, the situation for multi-wall tubes is less
clear from both theory and experiment. Many experiments find evidence
for diffusive electronic transport
\cite{langer,schoenenberger,bachtold_nat}, however, ballistic
transport has been also reported \cite{frank}. 

The origin of the electron scattering mechanism that is at work 
in multi-wall tubes, but obviously inefficient in single-wall tubes, is not 
well understood. It could be attributed to the typically larger
diameter of multi-wall tubes, which is accompanied by smaller
sub-band energies. As a consequence, higher sub-bands
become occupied by electrons or holes when the Fermi energy is
shifted off half filling due to doping or an external electrical
potential \cite{schoenenberger,schoenenberger2}.  
Unlike the two lowest bands, which are protected against backward scattering 
by a certain symmetry of the tube states \cite{nakanishi},  higher
bands are not. They are therefore more sensitive to impurity scattering.
It has been shown that this 
effect causes the unusually high resistivity of semi-conducting 
single-wall tubes \cite{mceuen}.

Another possible scattering mechanism specific for multi-wall tubes or
bundles is inter-tube Coulomb coupling. Since typically transport in
such systems is supported only by a fraction of tubes
\cite{langer,bachtold_nat,schoenenberger,frank}, 
Coulomb-force mediated scattering between electrons of active and
passive tubes can be a source of additional resistance as well. 

In this work we analyze and compare these two scattering
mechanisms. Our main result is that for multi-wall tubes with a typical 
amount of doping (as e.g.~in \cite{schoenenberger2}) backward
scattering at dopants exceeds by far the backscattering caused by 
inter-tube electron-electron interaction. 

In principle, inter-tube scattering can also be caused by
incommensurate tube structures. When adjacent tubes
have different molecular structure, electrons of one tube experience 
the static lattice potential of the other tube as 
incommensurate with the potential of the lattice of the
own tube. As a consequence, scattering occurs. A thorough analysis
of this effect, and a quantitative comparison with the aforementioned
scattering processes, however, is beyond the scope of the present
publication. Nevertheless, we would like to refer to recent works
Ref.\ \cite{kolmogorov} and Ref.\ \cite{roche} that addresses related
effects of incommensurabilities in multi-wall tubes.      

In the analysis presented below we neglect inter-tube tunneling.
We justify this by the fact that many experiments find evidence for
that the current in a multi-wall tube flows predominantly through the outer
tube \cite{langer,bachtold_nat,schoenenberger,frank}.  
There is also theoretical evidence that 
inter-tube tunneling might be strongly affected by
incommensurabilities  \cite{maarouf,roche}.

We begin by briefly reviewing the electronic structure of carbon
nanotubes. The following section \ref{sec-matrix} provides 
the matrix elements for scattering by impurities and by
electron-electron interaction. In section \ref{sec-comparison} we
evaluate and compare the results for real systems, and give a conclusion
in section \ref{sec-discussion}.

\section{Electronic structure}\label{sec-electronic}
A $(n_1, n_2)$-tube can be viewed as a 2D  graphite lattice 
that is bended in a way that a lattice vector $c_{n_1,n_2} = n_1 a_1 +
n_2 a_2$, where $a_1, a_2$ are primitive lattice vectors of length
$a=2.47\angst$, becomes a circumferential vector of a cylinder. 
Closing the tube periodically restricts the lattice
momentum $k$ to sub-bands defining lines $k \cdot c_{n_1,n_2} = 2 \pi l$ in
$k$-space, where the integer $l$ is the band index \cite{saito}.
Valence and conductance bands meet at the two so-called Dirac-points
$K_{\alpha = \pm 1}$. Tubes obeying $n_2-n_1 = 0\: {\rm mod}\: 3$ have
the two Dirac-points in the allowed $k$-space and are thus metallic,
while all other tubes are semi-conducting. 

In many experiments the Fermi-energy is significantly
off half-filling, either due to the influence of external
electrostatic potentials or doping
\cite{schoenenberger,schoenenberger2}. The shift of the Fermi-energy 
is usually less than the sub-band energy separation in single-wall
tubes. Hence, even in doped single-wall tubes only the lowest sub-band
($l=0$) is occupied. Multi-wall tubes, however,  have typically much
larger radii and therefore much smaller sub-band energies than
single-wall tubes. Consequently, the shift of the Fermi-energy 
usually leads to the occupation of higher sub-bands in multi-wall tubes
\cite{schoenenberger,schoenenberger2}.

The low energy physics of nanotubes is determined by electronic states in the
vicinity of the Dirac-points $K_{\alpha}$. Neglecting curvature
effects, their structure can be conveniently taken from the
corresponding electronic states of plane graphite:
With $A_\alpha(r), B_\alpha(r)$ two 
degenerate, orthonormal Bloch eigenstates of 2D graphite at $K_\alpha$,
near-by states with momentum $k=K_\alpha+q$ can be expanded as
\cite{slonczewski,divincenzo}  
\ben \label{eigenstate}
\psi_{\alpha q} = \fr{e^{iq \cdot r_\parallel}}{\sqrt 2}
\left(A_\alpha + f_q B_\alpha \right)   \:.
\een
We set $q= |q| ( \cos \vartheta, \sin \vartheta )$ with respect to a
fixed coordinate frame of choice. Then the relative phase is
$f_q =\pm \e^{i \alpha \vartheta}$ for valence band ($-$) and
conduction band ($+$).
In first order of $q$, the energy dispersion around $K$
is conical \cite{slonczewski}, $ E_{K+q} = \pm v_F |q|$,
with $ v_F \approx 5.4 eV \angst$ \cite{painter} (we use units in
which $\hbar \equiv 1$ and $k_B \equiv 1$).
The state $\psi_{\alpha q}$  can be viewed as a pseudo-spinor,
where the two pseudo-spin polarizations refer to Bloch eigenstates
$A_\alpha, B_\alpha$ \cite{divincenzo}.  
It is convenient to introduce a mixing angle $\gamma_{qq^\prime}$ that
measures the pseudo-spinor overlap of two states $\psi_{\alpha q}$ and
$\psi_{\alpha q^\prime}$ by
\be 
\cos \gamma_{qq^\prime} = \left| \av{ e^{-iq \cdot r_\parallel}  \psi_{\alpha q},
e^{-iq' \cdot r_\parallel} \psi_{\alpha q'}}\right|\:. 
\ee
Geometrically, $\gamma_{qq'}$ is half the angle enclosed by 
$q$ and $q'$. Eigenstates of opposite momentum have orthogonal
pseudo-spin polarizations, i.e. $\gamma_{q, -q}= \pi/2$. This is the 
reason for the strong suppression of backscattering in metallic
single-wall nanotubes \cite{nakanishi}. 

The two-dimensional Bloch states \Ref{eigenstate} translate into 1D 
tube-states $\psi_{\alpha q}^{(l)}$ of sub-band $l$ by 
\ben\label{tubestate}
\psi_{\alpha lk}^{(tube)} \equiv
\psi_{\alpha q}^{(gr)}\:, \quad q = (k,q_l)\:,
\een
where $k$ is the momentum along the tube axis (relative to $K_\alpha$)
and $q_l = 2\pi l / |c_{n_1,n_2}|$ is the transversal momentum (relative
to $K_\alpha$) of sub-band $l$.
The in-plane coordinates $x,y$  and the radial coordinate 
$z$  of the plane graphite-lattice coordinate frame thereby
become longitudinal ($x$), circumferential ($y$), and
local off-plane coordinate ($z$) in a curved tube coordinate frame.
The coordinate $y$ is $2\pi R$-periodic, where $ 2\pi R \equiv
|c_{n_1,n_2}|$, and parallel to $c_{n_1,n_2}$. 

\begin{figure}
  \begin{center}
    \epsfxsize=9.0cm
    \leavevmode
    \epsffile{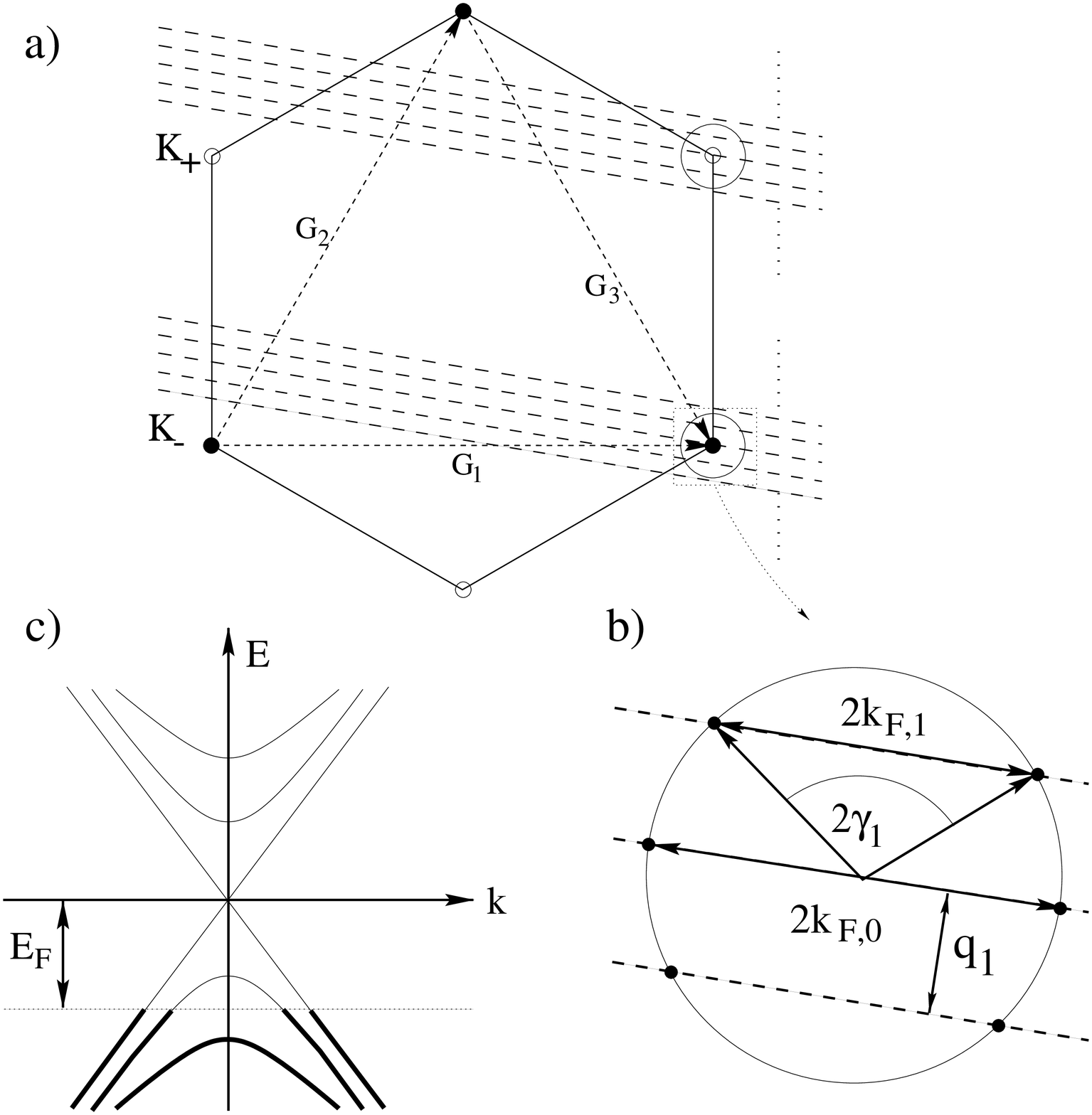}
    \caption{a) Brillouin zone of planar graphite with primitive
    reciprocal lattice vectors $G_1$, $G_2$, $G_3$, and Dirac-points
    $K_+$ and $K_-$. The dashed lines (to be continued over the entire
    zone) represent the allowed momentum states of a metallic
    $(45,15)$-carbon nanotube.  
    Their intersections with the lines $E=E_F$ (rings around the
    Dirac-points) define the Fermi-points of the tube sub-bands. b)
    Local structure in the vicinity of $K_-$. c) Energy dispersion of
    sub-bands in the vicinity of $K_-$.}
    \label{fig1}
  \end{center}
  \vspace{-0.5cm}
\end{figure}

The energy dispersion near $E_F$ at $K_-$ and $K_+$ of a sub-band $l$
results from the dispersion $E_{K+q} = \pm v_F |q|$
confined to the line $q_y = q_l$. Hence, electrons in the
lowest sub-band ($l=0$) are massless Dirac fermions, while electrons
in higher sub-bands ($l \neq 0$) acquire a mass $\Delta = v_F q_l$
(cf. figure \ref{fig1}).
Further, in the massive bands Fermi-point states 
at $-k_{F,l}$ and $k_{F,l}$ have no longer orthogonal pseudo-spin
polarizations,  as it is the case for a massless band, but
rather mix with a mixing angle 
$$
\gamma_{k_{F,l},-k_{F,l}} \equiv
\gamma_l = \arctan k/q_l < \pi/2\:.  
$$

\section{Matrix elements}\label{sec-matrix}
This section provides the matrix elements for intra-sub-band backward
scattering caused by interaction with dopants, and by interaction with
electrons in adjacent tubes.

\subsection{Impurity scattering}\label{subsec-impurity}
Recently, it has been observed \cite{schoenenberger2} that multi-wall
tubes in air are substantially hole-doped. The measured shift of the
Fermi-energy of $\Delta E \sim -0.3eV$ indicates a dopant-charge
concentration of about one elementary charge per 500 carbon atoms
\cite{schoenenberger2}. In  view of this rather high concentration,
we consider dopants as the main source for inelastic scattering. 

Earlier publications focused on substitutional disorder and lattice defects
\cite{chico,tamura,white_todorov,anatram,kostyrko}. 
While the potentials created by these lattice imperfections vary rapidly on 
a scale of order or less than the lattice constant $a$, it is likely
that the effective dopant potential  
$V_d(x,y)$ on the tube-surface is rather smooth on that scale. The
reason for that being that the dopant might be located in a
distance $b \gapprox a$ from the tube surface and additionally may have a spatial
extension. In the following we therefore assume that the dopant
potential is smooth. 

In this case scattering of electrons from one Dirac-point to the other
is strongly suppressed \cite{nakanishi}, such that we can confine our
considerations  to scattering events within states in the vicinity of
one Dirac-point.  Following Ando et al. \cite{nakanishi}, and making use of the 
$k\cdot p$-approximation \cite{slonczewski,divincenzo} we obtain
for intra-sub-band backscattering  the matrix element
\bean
M_l^{(i)}  &=& \int d^3r \; \psi_{\alpha l -k_{F,l}}^*(r) V_d(r) \psi_{\alpha
  l k_{F,l}}(r) \nn \\
&=& L\inv \cos \:\gamma_l \:\hat U_d(2k_{F,l})\:, \label{dop-matrix}
\eean
where $L$ is the length of the tube, and $\hat U_d$ the Fourier
transform of the effective 1D potential 
\be
U_d(x) = (2 \pi R)\inv \int_0^{2 \pi R}
V_d(x,y) dy\:. 
\ee

Independent of the precise form of the dopant potential 
the following observations hold \cite{nakanishi}:
For a massless band ($l=0$) the matrix element vanishes, since 
$\cos \gamma_{l=0} =0$ due to the orthogonality of pseudo-spin
polarizations of states at $-k_{F,0}$ and $k_{F,0}$. For higher
sub-bands this is not the case, and 
$M_l^{(i)}$ can assume appreciable values \cite{mceuen}, depending on
the mixing angle $\gamma_l$.  
We emphasize that typically $2 k_{F,l} \ll  \sim a\inv$, 
which means that the backscattering coupling  $M_l^{(i)} \propto \hat
U(2k_{F,l})$ is {\em not} suppressed  by a large transfered momentum.

For the purpose of quantitative estimates we need to further specify 
the dopant potential. Modelling the dopant as an elementary charge $e$
located in a distance $b$ to the surface of the tube, its regularized
Coulomb-potential on the tube may be written as
\ben\label{dop-pot}
V_d(x,y) = \fr{e^2}{ \sqrt{x^2 + S_{bc}(y/R)^2} },
\een
where $ S_{bc}(\fie)^2 = c^2 + b^2 + 4R(b+R) \sin^2 (\fie/2) $. The 
length $c$ of order $1 \angst$ takes into account the finite
width of the graphite layer as well as the spatial extension of the dopant
charge. From this potential we obtain 
\ben\label{dop_u}
\hat U(q; b,c) = 2 e^2 \int_0^{2 \pi } \fr{d\fie}{2\pi } K_0(q\:S_{bc}(
\fie))  
\een
where  $K_0$ is the modified Bessel-function of the second kind. 
The smoothness condition requires $(b^2+c^2)^{1/2} \gapprox a$.

The exact values of the parameters $b$ and $c$ are hard to
determine. Fortunately, it will turn out that the
dependence on these parameters is relatively week for
$(b^2+c^2)^{1/2}$ being in a rather wide regime $\approx 2.0 \cdots 10 \angst$
(Sec.\ \ref{sec-comparison}, Fig.\ \ref{fig2}).
The estimates given below seem to be not very sensitive to
the details of the dopant-potential, which also motivates our specific choice 
\Ref{dop-pot}. 

\subsection{Electron-Electron scattering}\label{subsec-electron}
The distance between adjacent walls in a multi-wall tube is as like in 
graphite approximately $d=3.4\angst$. Because of this relatively large 
separation, we assume that the inter-tube electron-electron
interaction potential can also be viewed as a smooth
potential. Consequently, we will again neglect scattering 
transitions where electrons change from $K_-$ to $K_+$, and calculate
the matrix elements for the remaining backscattering events in 
a similar way as for the impurity scattering, as it is briefly
outlined in the following.

The matrix element for inter-tube electron-electron backward
scattering is
\bea
M^{(e)}_{ll'} &=&  \int d^3 r d^3 r' 
\psi_{\alpha l k_3}^*(r) \psi_{\alpha lk_1}(r) V(r,r')  \\
& & \quad \times \psi_{\alpha' l'k_4}^*(r') \psi_{\alpha' l' k_2}(r')\:,
\eea
where $k_1=-k_3 \approx k_{F,l}$, $k_2= -k_4 \approx -k_{F,l'}$, and $V(r,r')$ is
the interaction potential as a function of the respective tube
coordinates. 
Using Eq.s \Ref{tubestate} and \Ref{eigenstate} 
and neglecting integrals that contain mixed terms
$A^*(r)B(r)$, the matrix element becomes
\bea
&\fr{1}{4}& \int d^3 r d^3 r' (\rho_A+f^*_3 f_1 \rho_B
)V \times \\
& & \quad ( \rho_A' + f^*_4 f_2 \rho_B' ) \e^{i(k_1-k_3)x + i(k_2-k_4)x'}\:.
\eea
Here, $\rho_{A/B}$, $\rho_{A/B}'$ denote the densities of
eigenstates $A$ and $B$ on the two tubes.
Since the transfered momentum $k_1 - k_3 = k_2 - k_4 $ is
small compared to $a\inv$, the microscopic structure of the
densities $\rho_A, \rho_B $ is unimportant. This allows us to
approximate $\rho_{A/B}$ by a homogeneous density on the
tube surface,
$ 
\rho_A = \rho_B = \delta(z)/(2\pi R L )
$, and so for $\rho_{A/B}'$.
In this approximation, the matrix element for backward scattering in a
tube of length $L$ is
\ben\label{matrix_element}
M_{ll'}^{(e)} = L\inv\delta_{k_1-k_3,k_2-k_4} \cos \gamma_l \cos \gamma_{l'}
\:\hat U_e(k_1 - k_3)\:,
\een
where $\hat U_e$ is the Fourier transform of the 
effective 1D electron-electron interaction potential 
\be
U_e(x-x') = \int_0^{2\pi R} \fr{dy}{2\pi R}
\int_0^{2\pi R'} \fr{dy'}{2 \pi R'}
 V (x-x', y, y')\:.
\ee

The factor $\cos \gamma_{l} \cos \gamma_{l'}$ in the matrix element
\Ref{matrix_element} indicates the same characteristic suppression by
orthogonal pseudo-spin polarizations as we have seen for the
backscattering by dopants, Eq. \Ref{dop-matrix}. Thus, also the
backscattering by electron-electron interaction vanishes for metallic
bands, independently on the particular form of
the interaction $V(r,r')$.

We describe the interaction between electrons on coaxial tubes of radii $R$
and $R'$ by the regularized Coulomb potential
\be
V(x-x',y,y') = \fr{e^2}{\sqrt{ (x-x')^2 + 
S_{RR'\tilde c}^2(\fr{y}{R}-\fr{y'}{R'})}}\:,
\ee
where
\be
 S^2_{RR'\tilde c}(\fie) = \tilde c^2 + (R-R')^2 + 4RR'
 \sin^2(\fie/2)\:.
\ee
The parameter $\tilde c \sim 1\angst $ reflects the extension of
the electron densities in the radial direction.
For this interaction we obtain
\ben\label{ee-mat}
\hat U_e(q; R,R',\tilde c) = 2 e^2 \int_0^{2\pi} \fr{d\fie}{2\pi} 
K_0(q S_{RR'\tilde c}(\fie))\:.
\een
The dependence of this Fourier coefficient on $\tilde c$ is similarly
week as like the dependence of $\hat U_d(q;b,c)$ on $c$.  

\section{Comparison}\label{sec-comparison}
To be specific, we take parameters that are typical for the recent
experiment on multi-wall nanotubes by Sch\"onenberger et
al. \cite{schoenenberger}: $D= 10nm$ for the diameter of the outer 
tube, and a Fermi energy $E_F = -0.3\;eV$ (relative to the energy of the 
Dirac-point states). Assuming a conical dispersion $E_{K_\pm+q} = v_F
|q|$ with $v_F= 5.4 eV \angst$, it follows that a total of ${\cal N}=10$
spin-degenerate sub-bands are occupied. We label the five sub-bands at
each Dirac-point by $l = 0, \pm 1, \pm 2$. The corresponding 
mixing angles are given by $ \cos \gamma_l = .36\, |l|$,
the $1D$-Fermi momenta are  $k_{F,0} = .056 \,\angst\inv$, $k_{F,\pm1} =
.052\, \angst\inv$ and  $k_{F,\pm 2} = \,.039 \angst\inv $.

The total density $n_i$ of the  dopant charges close to the tube
surface can be deduced from charge neutrality 
\cite{schoenenberger2}: 
the total density of electrons $n_e$ that is expelled from the tube in
its neutral state (where $E_F = 0$) must equal the density $n_i$ of 
dopants. For the chosen parameters we find $n_e=n_i = 3.0 nm\inv$.

The efficiency of the two scattering mechanisms under 
discussion cannot be directly compared by their matrix elements
presented in the previous section. A quantity that is suited
for a comparison is for example the transport scattering time $\tau$.
To proceed with a reasonable amount of effort,  we calculate $\tau$
within the scope of a Boltzmann-equation approach \cite{abrikosov},
where we restrict ourself to inter-sub-band scattering only.  

\begin{figure}
  \begin{center}
    \epsfxsize=8.0cm
    \leavevmode
    \epsffile{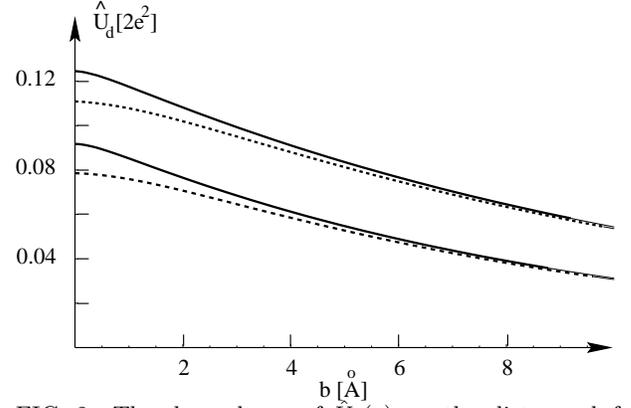}
    \caption{The dependence of $\hat U_d(q)$ on the distance $b$
    for parameter $c=0.5\angst$ (dashed) and $c=2.0\angst$ (solid). The
    wavevectors are choosen to be $q = 2 k_{F,1}$ for the lower, and
    $q= 2 k_{F,2}$ for the upper curves. }
    \label{fig2}
  \end{center}
\end{figure}

For scattering at dopants we find in this way
\bean\label{impurity-tau}
\fr{1}{\tau^{(i)}_l} &=& \fr{2 n_i}{v_{F,l}} |L M_l^{(i)}|^2 \\
&=& \fr{2 n_i}{v_{F,l}} \cos^2 \gamma_{l} \: |\hat U( 2k_{F,l}; b,c)|^2\:.
\eean
The dependence of $\hat U(q;b,c)$ (Eq.\Ref{dop_u}) on the parameters $b$
and $c$ is rather weak, as shown in Fig.\ \ref{fig2}. For transfered
momenta $q= 2 k_{F,1/2}$ the potential $\hat U$ varies with $b$ and
$c$ ranging from $2.0 \cdots 10 \angst$ and $0.5 \cdots 2 \angst$ 
by less than a factor of 3. 
The transport scattering times in this range of parameters are
\bean\label{tau_imp}
\fr{1}{\tau_{0}^{(i)}} =0, \quad 
\fr{1}{\tau_{1}^{(i)}} =.012\cdots 0.076\,eV, \nn \\
\fr{1}{\tau_{2}^{(i)}} =.20 \cdots 0.82\,eV\:, 
\eean
where the lower values belongs to $b=10\angst$ and $c=2.0 \angst$, 
the higher to $b=2.0\angst$ and $c=0.5\angst $. The 
corresponding mean free paths, formally defined by $ l_l \equiv
v_{F,l} \tau^{(i)}$, are
$l_0 = \infty$, $l_1 = 0.66\cdots 41 nm$, and $l_2 = .46 \cdots 1.9  nm$.

The rather large
values of the inverse transport times for $l=1$ and 2 
indicate that as soon as massive bands are involved,
backscattering at dopants indeed gives rise to a significant intrinsic
resistivity. Taking the calculated mean free paths $l_1$ and $l_2$
literally would even result in a higher resistivity than is actually 
observed \cite{bachtold_nat,schoenenberger}. The reason for this
overestimation could be an improper modelling of the dopant impurities
or the neglect of screening.

For the transport time caused by backscattering
of electrons in sub-bands $l$ and $l'$ of different tubes we
obtain within the Boltzmann-equation approach
\bean\label{ee-tau}
\fr{1}{\tau_{ll'}^{(e)}} &=& \fr{ T}{2 \pi v_{F,l} v_{F,l'} } |L
M_{ll'}^{(e)}|^2\\
&=& \fr{T}{2\pi v_{F,l} v_{F,l'}} \cos^2 \gamma_l \cos^2 \gamma_{l'}
|\hat U_e(2k_{F,l};R,R',\tilde c) |^2 \nn \:.
\eean
Here it is assumed that the Fermi-momenta in the participating channels 
match, $k_{F,l}=k_{F,l'}$, otherwise the scattering rate is strongly
suppressed (see Eq. \Ref{matrix_element}).

Evaluating Eq.\ \Ref{ee-tau} for $R=50\angst$, $R'=R-3.4\angst$, and 
$\tilde c = 1.0\angst$, we obtain in this case the transport scattering
times 
\ben\label{tau_ee}
\fr{1}{\tau_{00}^{(e)}} = 0, \quad
\fr{1}{\tau_{11}^{(e)}} = 4.3\: 10^{-4}  T\:, \quad 
\fr{1}{\tau_{22}^{(e)}} = .028 \, T \:.
\een

Even at room temperature these inverse transport times are by orders of
magnitude smaller than those caused by dopant scattering. This
strong suppression is mainly due to the smallness of the 
dimensionless parameter $ T\,/\, v_F n_i \sim \pi T\, / \, {\cal N}
|E_F|$. (The matrix elements $M_{ll}^{(e)}$ and $M_{l}^{(i)}$ are of
comparable size.) 
Interpreting the inverse thermal wavelength as the density
$n_T$ of thermally activated electrons/holes, $n_T = T / v_F$, the
small value of $\tau^{(i)}/\tau^{(e)}$  corresponds to the fact that
for the considered parameters the density $n_i$ exceeds $n_T$ by a
large factor $ {\cal N} |E_F|\,/ \, \pi T$.

\section{Discussion}\label{sec-discussion}
The preceding estimates show that under typical experimental
conditions scattering by dopants can be a source of
significant intrinsic resistance in multi-wall nanotubes. 

The typically larger diameter of multi-wall
tubes entails the  occupation of higher sub-bands, which are, in
contrast to the massless sub-bands ($l=0$), no longer protected
against backscattering by orthogonal pseudo-spin polarizations of 
states at opposite Fermi-points. 
The same effect explains \cite{mceuen} the high resistivity of gated
semi-conducting single-wall nanotubes, which naively could be expected
to be as well conducting as metallic single-wall tubes. In fact, it
has been already speculated in  Ref.\ \cite{mceuen} 
that the resistance of multi-wall tubes could have the same origin.

The conclusion that the enhancement of backscattering in
multi-wall tubes is due to their larger radii is  not at odds
with the results of White and Todorov \cite{white_todorov}. 
Their observation that impurity scattering decreases with increasing
diameter of tube applies for scattering within the {\em massless}
bands, while our conclusion relies on the investigation of backscattering in the
{\em massive} bands. 

For Coulomb interaction with electrons in inner tubes we observe
a qualitatively similar behaviour: the suppression of
backscattering in the massless bands due to anti-symmetry is 
suspended in the massive bands. Quantitatively, we find however that
for a typicall amount of doping \cite{schoenenberger2} the
backscattering rate caused by  intra-tube electron-electron
interaction is by orders of magnitudes smaller than the rate caused by
the interaction with dopants. 

Combining these two results, on might say that the observed
non-ballistic electronic transport in multi-wall tubes 
is primarily due to the enhanced backscattering at dopant impurities, and not an
effect of interactions between different shells. 

The reported ballistic transport in multi-wall tubes in the experiment
by Frank et al. \cite{frank} does not contradict the picture presented
here. Differing from the others, in this experiment the tubes have
been contacted by partially immersing them into liquid mercury.
Thereby the tubes may have been cleaned from surface impurities and
may have been also protected  from absorbing surface dopants \cite{frank}. 
For this reason, in this experiment the Fermi-energy may be
close to the energy of the Dirac points, such that only the massless
bands are occupied (for which backscattering by impurities or electrons
in other shells is suppressed).  Or, even when higher bands are
occupied, due to the absence of surface impurities backscattering is
insignificant. 

The intra-sub-band scattering rate has been considered
as an indicator for the strength of two certain 
scattering mechanisms. For a more quantitative comparison with
experimental results it is necessary to include also 
(back)scattering between different sub-bands.
Further improvements might be achieved by a more precise modelling 
of the dopant potential, and taking into account effects
of screening and  electronic correlations. 

Finally,  we like to stress that the present work focused 
on impurity- and electron-electron scattering only. For a 
complete picture of the transport in nanotubes 
it is necessary to investigate other possible scattering mechanisms. 
Particularly, further investigations of the effects of lattice
incommensurabilities on transport are desirable.

I thank S. Ernst for critically reading the manuscript.

\end{document}